\documentclass[final,english]{bullsrsl}[2022/06/15]

\usepackage[latin1]{inputenc}
\usepackage[T1]{fontenc}

\usepackage{natbib} 
\usepackage{graphicx}

\begin{document}
\title{TESS Exploration of Targets Investigated for the Nainital-Cape Survey Project}

\author[affil={1,2},corresponding]{Athul}{Dileep}
\author[affil={1}]{Santosh}{Joshi}
\author[affil={3}]{Donald}{Wayne Kurtz}
\affiliation[1]{Aryabhatta Research Institute of Observational Sciences, Nainital 263001, Uttarakhand, India }
\affiliation[2]{Mahatma Jyotiba Phule Rohilkhand University, Bareilly, Uttar Pradesh 243006, India }
\affiliation[3]{Centre for Space Research, Physics Department, North-West University, Mahikeng 2735, South Africa}
\correspondance{dileep@aries.res.in}
\date{13th April 2023}
\maketitle

\begin{abstract}
The Nainital-Cape Survey was initiated more than two decades ago aiming to search for and study the pulsational variability in two subclasses of chemically peculiar (CP) stars, namely the Ap and Am stars.
In this paper, we present the TESS photometry of 4 targets out of the 369 sample stars observed under the survey, which were not studied before using TESS data. Our results suggest that HD\,34060 is a rotational variable, HD\,25487 is of eclipsing nature, HD\,15550 exhibits pulsational variability while HD\,48953 is a non-variable star. The diverse variability detected in the studied sources places important constraints for the study of the internal structure and evolution of the CP stars in the presence of surface inhomogeneity, magnetic field, rotation and pulsation.

\end{abstract}


\section{Introduction}
The ``Nainital-Cape'' (N-C) survey is one of the largest ground-based surveys of photometric variability for Chemically Peculiar (CP) stars in terms of time span and number of samples monitored. The survey was conducted using the 1.04-m Sampurnanand telescope of the Aryabhatta Research Institute of Observational Sciences (ARIES; Nainital, India) and the 0.5-m telescope at the Sutherland observing site of the South African Astronomical Observatory (SAAO; Sutherland, South Africa). Under this survey, a total of 369 stars were monitored aiming to search for and study the pulsational variability in CP stars. These stars are designated `CP' because their spectra exhibit abundance anomalies that manifest as extreme over- or under-abundances of various chemical elements. The pulsations in CP stars were reviewed recently by \cite{doi:10.1146/annurev-astro-052920-094232}; the reader is referred to this  article and references therein for an in-depth knowledge of the subject. The primary target selection criteria, adopted observing strategy, and results obtained from the N-C legacy survey are summarized by \cite{2000ashoka}, \cite{2001martinez}, and \cite{joshi2003,2006joshi,joshi2009,joshi2010,joshi2012,joshi2016,joshi2017,joshi2022}. 

\begin{table}[!ht]
\centering
\caption{Variability types given by \cite{balona} for the sample observed under the N-C survey. The classes which were uncertain are flagged with `:' and the `+' symbol refers to stars showing more than one type of variability. The abbreviation used for each class is described in the footnote of this table.}
\label{Tab:Table 1}
\bigskip
\begin{tabular}{cccccc}
\hline
\textbf{Class} & \textbf{Number} & \textbf{Class} & \textbf{Number} &\textbf{Class} & \textbf{Number} \\
\hline
ACV        & 129 & SXARI+MRP      &   4 & ACV:+roAp &   1 \\
SXARI      &  71 & SPB            &   3 & DSCTH+ACV &   1 \\
DSCT       &  21 & DSCTH          &   2 & ACV+MRP   &   1 \\
ACV+roAp   &  17 & EB             &   2 & ACYG      &   1 \\
ROT        &  15 & SPB:           &   2 & SXARI:    &   1 \\
roAp       &  11 & DSCTU+ACV+ROAP &   2 & ROT+MRP   &   1 \\
EA         &   7 & ACV:+FLARE     &   1 & ZAND      &   1 \\
DSCT+ACV   &   5 & ACV+FLARE      &   1 & BE+ROT    &   1 \\
GDOR       &   5 & DSCTH+EA       &   1 & SPB+SXARI &   1 \\
MAIA+SXARI &   4 & ACV:           &   1 & SORI+ROT  &   1 \\
           &     &                &     & NV        &  51 \\
\hline
\end{tabular}
ACV-$\alpha^{2}$ CVn star; SXARI-SX Ari star; DSCT-$\delta$ Scuti star; roAp-rapidly oscillating Ap star; ROT-Hot stars with rotational modulation; EA-Algol-type eclipsing system; EB-$\beta$ Lyrae-type eclipsing system; GDOR-$\gamma$ Doradus star; MAIA-Maia variable; MRP-star emitting radio pulses; SPB-Slowly pulsating B star; ACYG-$\alpha$ Cygni star; ZAND-Symbiotic variable of the Z Andromedae type; BE-Be star; SORI-$\sigma^{2}$ Orionis star; FLARE-Flare events; DSCTH-$\delta$ Scuti star with a frequency higher than 50 $d^{-1}$; DSCTU-$\delta$ Scuti star with a frequency higher than 60 $d^{-1}$; NV-non-variable star.
\end{table}

Of the 369 stars observed under this survey, 365 are classified by \cite{balona} via analysis of Transiting Exoplanet Survey Satellite (TESS) time-series data, the results of which are tabulated in Table \ref{Tab:Table 1} where 51 stars are classified as non-variable. Here, we present the TESS photometry of the 4 remaining stars that are missing in the catalogue of \cite{balona}. The paper is organised as follows. The photometric observations and data analysis technique are discussed in Section \ref{obs}. The results obtained from our analysis are summarised in Section \ref{indi}. Finally, the future prospects of the ongoing project are outlined in Section \ref{future}.

\section{Observations and Data Analysis}\label{obs}

TESS is an all-sky survey telescope whose prime objective is to detect exoplanets orbiting nearby bright stars using the transit method \citep{2014JAVSO..42..234R}. Apart from its prime mission, it also provides time-series photometry for asteroseismology of stars brighter than 12th magnitude. TESS Target Pixel Files (TPF) and Light Curve Files are available with a cadence of 20 and 120 seconds (short cadences) while Full-Frame Images (FFIs) are taken every 30 minutes for TESS cycles 1 and 2, 10 minutes for TESS cycles 3 and 4, and 200 seconds from TESS cycle 5 onwards (long cadences). The data products can be downloaded from the Mikulski Archive for Space Telescopes (MAST). TESS Light Curve File objects contain two kinds of flux: the Simple Aperture Photometry (SAP) flux and the Pre-search Data Conditioning SAP (PDCSAP) flux. For the PDCSAP flux, long-term trends have been removed from the data using Co-trending Basis Vectors (CBVs). For our purpose, we have used 120-s cadence data downloaded from MAST using the python package \emph{lightkurve} (\citealt{2018ascl.soft12013L}). Almost all the targets monitored under the N-C survey project have been classified by \cite{balona} except four stars, namely, HD\,34060, HD\,48953, HD\,25487, and HD\,15550. Therefore, the PDCSAP flux of these stars were downloaded from MAST.

The cleaned light curves were subjected to frequency analysis to detect and identify the dominant frequencies using the $Lomb-Scargle$ method \citep{1982ApJ...263..835S} implemented in the \mbox{PERIOD04} package developed by \cite{2014ascl.soft07009L,2005CoAst.146...53L}. A signal--to-noise ratio (SNR) above 4 is considered as a criterion for selecting significant frqeuencies in ground-based observations. For space-based observations, this SNR limit underestimates statistical noise fluctuations introduced by the large number of data points spanning several months to years. \cite{2016A&A...585A..22Z} showed that a SNR limit of 5.6 is a safe criterium to avoid the identification of false peaks. We therefore adopted it for this study.

\begin{figure}[!ht]
\centering
\includegraphics[width=14cm]{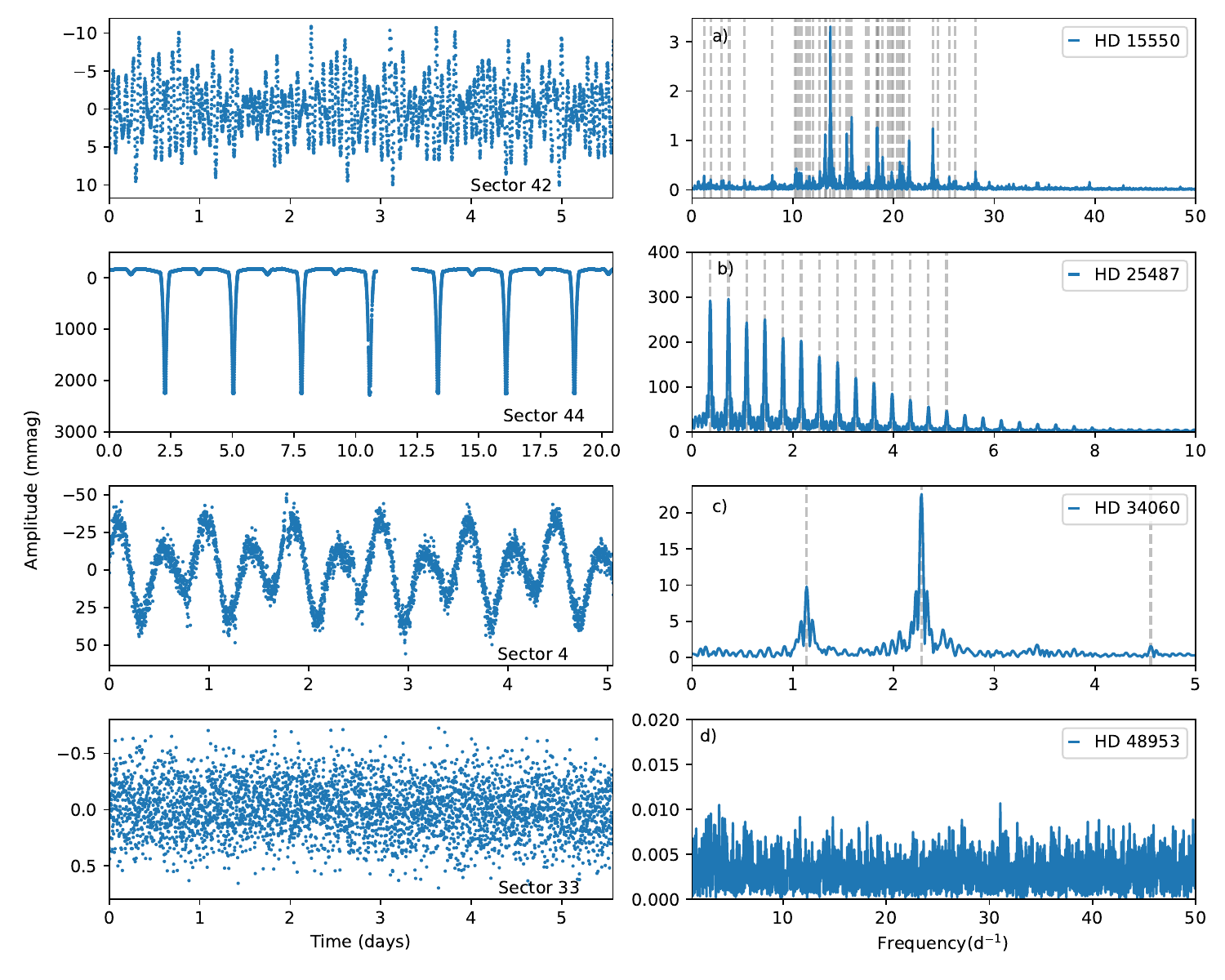}
\bigskip
\begin{minipage}{12cm}
\caption{The TESS light curves (left panels) and corresponding amplitude spectra (right panels) of HD\,15550, HD\,25487, HD\,34060, and HD\,48953 obtained from sectors 42, 44, 4, and 33 respectively. The significant frequencies are indicated with vertical gray dashed lines. The harmonics of the fundamental frequency are clearly seen in the amplitude spectrum of the eclipsing (HD\,25487) and rotational (HD\,34060) variables.}
\label{fig:lc}
\end{minipage}
\end{figure}

The light curves and corresponding amplitude spectra of the given sectors  for the studied targets are shown in the left and right panels of Fig.\ref{fig:lc}, respectively. Using the multi-component sinusoidal model, we derived the frequencies, amplitudes and SNR, which are listed in Table \ref{Tab:Table 2}.

\section{Results and Discussion}\label{indi}

The results inferred from our study are summarized in Table \ref{Tab:Table 2}. The classification was done according to the general scheme of variable star classification (e.g. visual inspection of shape of the light curves and position of peaks in the frequency spectra).  The following sub-sections provide brief summaries of the information on individual stars available in the literature, along with our results from the present study.

\subsection{HD\,15550}

HD\,15550 (UU Ari) is classified as a non-radial $\delta$ Scuti ($\delta$\,Sct) pulsating variable with a period and $V$-band light variation of 0.068 d and 0.01 mag, respectively \citep{2000A&AS..144..469R}. 
Panel a) of Fig.\ref{fig:lc} illustrates the presence of multiple peaks in the frequency range 5-50\,d$^{-1}$. We found 41 new significant frequencies, of which the first three are listed in Table\,\ref{Tab:Table 2}. The presence of light modulation seen in the light curve is another clear evidence of the multi-periodic nature of this star. However, the frequencies reported by \cite{2000A&AS..144..469R} are not in agreement with the frequencies detected in the TESS data. Hence, this star needs to be revisited for a detailed asteroseismic study.

\subsection{HD\,25487}

HD\,25487 is a well known eclipsing binary system consisting of a B8Ve primary and another cooler K0IV companion \citep{1959PASP...71..345A}. The primary component was classified as a classical Be (Be) star by \cite{1982IAUS...98..261J} and recently studied using low-resolution spectroscopy by \cite{2021MNRAS.500.3926B}. It is to be noted that binarity is not uncommon in the case of Be stars (e.g. \citealt{2022ApJ...933L..34B, 2021AAS...23734904W, 2018ApJ...853..156W}). In another study, \cite{1993PASP..105..731V} put forward evidence of a B8V+K0IV system with an orbital period of 2.7688\,d. They suggested that the H$_\alpha$ emission (see Section 7 of their paper) is likely due to the presence of an accretion disk where mass is being transferred from the K-type to the B-type component. However, many Be stars also exhibit variability in H$_\alpha$ line profiles (e.g. \citealt{2022JApA...43..102B, 2013A&ARv..21...69R}). So we checked the photometric variability of this star using the TESS data. The PDCSAP flux for this star has extra noise introduced by the pipeline, an issue reported recently by many authors (e.g. \citealt{Holdsworth2021}; \citealt{kochukhov2022}). So, we used the SAP flux in this case for our analysis. The light curve and corresponding amplitude spectrum of HD\,25487 are shown in the  panel b) of Fig.\ref{fig:lc} which resembles the light curve of a typical detached eclipsing binary, which is in agreement with several previous studies. The primary eclipses have a flat bottom suggesting that the cooler star is larger in radius.
The  derived orbital period of this star is 2.7686\,$\pm$\,0.0038\,d. It is fully consistent with the value ($P_{orb}$\,=\,2.7688\,d) reported by \cite{1993PASP..105..731V}. A total of 14 orbital harmonics are still significant.
The presence of pulsation in either component (if any) can be only confirmed by removing the primary and secondary eclipses. 

\begin{table}
\centering
\begin{minipage}{13.3cm}
  \caption{The name of the star, sector, flux used, frequency, amplitude and SNR obtained for the targets under study. The last column gives the variability class assigned to the star from our study. The values in parentheses give the 1$\sigma$ uncertainty in the previous digit.
We use the same abbreviations for the variability types (Class) as in Table\,\ref{Tab:Table 1}.    
}
\label{Tab:Table 2}
\end{minipage}
\bigskip

\begin{tabular}{ccccccc}
\hline
\textbf{Star} & \textbf{Sector} & \textbf{Flux}&  \textbf{Frequency}  &\textbf{Amplitude} & \textbf{SNR} & \textbf{Class} \\
HD &  & \textbf{used} &(d$^{-1}$)  &(mmag) & (rad) & \\
\hline

15550&42&PDCSAP&13.7297(8)   &    3.3(1)& 25& DSCT \\
&&&15.842(2)   &    1.5(1)& 14& \\
&&&23.900(2)   &    1.3(1)& 26& \\
 
25487&44&SAP& $f_{orb}$\,=\,0.36122(9)   &    277(1)& 306& EA\\

34060&4&SAP &$f_{rot}$\,=\,2.2761(1)   &    22.9(1)& 36& ACV \\

48953 & 33 &PDCSAP& -  & -& - & NV\\

\hline
\end{tabular}
\end{table}

\subsection{HD 34060}

HD\,34060 was observed by \cite{1994MNRAS.271..129M} under the N-C survey, who reported a null result for the discovery of pulsations. Since the PDCSAP flux for this star was not available in the MAST archive, we performed the CBV correction in the SAP flux to obtain the corrected light curve. This star is a hot peculiar star of spectral type B9V(pSiCr) \citep{1978mcts.book.....H}. Based on the shape of the light curve, period of variability, and the presence of harmonics and subharmonics of the fundamental frequency ($f_1$\,=\,2.2761\,$\pm$\,0.0001\,d$^{-1}$, $f_2$\,=\,1.1378\,$\pm$\,0.0003\,d$^{-1}$ = $f_1/2$,  $f_3$\,=\,4.553\,$\pm$\,0.002\,d$^{-1}$ = $2f_1$, as shown in panel c) of Fig.\,1), we classify HD\,15550 as a rotation variable with a rotational frequency $f_{rot}$\,=\,2.2761\,$\pm$\,0.0001\,d$^{-1}$.

\subsection{HD 48953}
HD\,48953 is a cool star, classified as a magnetic (CP2) peculiar star by \cite {2012MNRAS.420..757W}. Based on the analysis of time-series data obtained with the {\it STEREO} satellites, these authors classified this star as a variable of period 2.8939\,$\pm$\,0.0024\,d. The PDCSAP light curve shown in the panel d) of Fig.\ref{fig:lc} does not exhibit any significant light variation up to the 3-$\sigma$ level ($\sigma$ $\sim$ 0.23 mmag). Therefore, we classify this star as a non-variable.

\section{FUTURE PROSPECTS}
\label{future}

In the present era, machine learning (ML) and artificial intelligence (AI) are promising tools to handle big data sets of space missions like {\it Kepler} and TESS. Application of these approaches is planned for the automatic classification of the variable stars and different types of transient sources. 
High-resolution spectra of one of the southern CP stars (HD\,100357) observed under the N-C survey were acquired with the Southern African Large Telescope (SALT). Their analysis is in progress for the determination of its basic parameters and chemical composition of its stellar atmosphere. The combined photometric and spectroscopic analysis of this star will provide the required input for a detailed study.   

\begin{acknowledgments}
We are grateful to the Indian and Belgian funding agencies DST (DST/INT/Belg/P-09/2017) and BELSPO (BL/33/IN12) for financial support receive to organize the third BINA workshop and other BINA activities. SJ acknowledges the financial support received from the BRICS grant DST/IC/BRICS/2023/5. AD acknowledges the financial support received from DST-INSPIRE Fellowship Programme (DST/INSPIRE Fellowship/2020/IF200245). This paper includes data collected by the TESS mission, obtained from the Barbara A. Mikulski Archive for Space Telescopes (MAST) at the Space Telescope Science Institute (STScI) and SIMBAD databases operated by the NASA Explorer Program and CDS, Strasbourg, France, respectively.
\end{acknowledgments}

\begin{furtherinformation}

\begin{authorcontributions}
This work is part of a long term survey program and collective efforts were made by all the co-authors with the relevant contributions.
\end{authorcontributions}

\begin{conflictsofinterest}
The authors declare no conflict of interest.
\end{conflictsofinterest}

\end{furtherinformation}

\bibliographystyle{bullsrsl-en}

\bibliography{S04-P03_DileepA}

\end{document}